\documentclass[10pt,twocolumn,superscriptaddress,aps,prd]{revtex4-1}
\usepackage{graphicx}
\usepackage{subfigure}
\usepackage{dcolumn}\usepackage{bm}
\usepackage[usenames]{color}
\usepackage{amsmath}
\usepackage{amssymb}
\usepackage{url} 
\usepackage{hyperref}

\def\lsim{\mathrel{\rlap{\lower4pt\hbox{\hskip1pt$\sim$}}
   \raise1pt\hbox{$<$}}}
\def\gsim{\mathrel{\rlap{\lower4pt\hbox{\hskip1pt$\sim$}}
   \raise1pt\hbox{$>$}}}

\begin{document}
\topmargin 0.0001cm
\title{Dark photon production through positron annihilation in
beam-dump experiments}  

\newcommand*{\INFNGE}{Istituto Nazionale di Fisica Nucleare, Sezione di Genova, 16146 Genova, Italy}
\newcommand*{\UNIGE}{Universit\'a degli studi di Genova, 16126 Genova, Italy}
\newcommand*{\INFNCT}{Istituto Nazionale di Fisica Nucleare, Sezione di Catania, 95125 Catania, Italy}
\newcommand*{\UDEA}{Universidad de Antioquia, Instituto de F\'isica, Calle 70 No. 52-21, Medell\'{i}n, Colombia}
\newcommand*{\INFNLNF}{Istituto Nazionale di Fisica Nucleare, Laboratori Nazionali di Frascati, C.P. 13, 00044 Frascati, Italy}
\newcommand*{\INFNRM}{Istituto Nazionale di Fisica Nucleare, Sezione di Roma, 00185 Roma, Italy}
\newcommand*{\UNIRM}{Universit\'a degli studi Roma La Sapienza, 00185 Roma, Italy}
\newcommand*{\Apr}{{A^\prime}}

\author {L.~Marsicano} 
\affiliation{\INFNGE}
\affiliation{\UNIGE}
\author {M.~Battaglieri} 
\affiliation{\INFNGE}
\author {M.~Bond\'i} 
\affiliation{\INFNCT}
\author{C.~D.~R.~Carvajal}
\affiliation{\UDEA}
\author {A.~Celentano} 
\affiliation{\INFNGE}
\author {M.~De~Napoli} 
\affiliation{\INFNCT}
\author {R.~De~Vita} 
\affiliation{\INFNGE}
\author {E.~Nardi} 
\affiliation{\INFNLNF}
\author {M.~Raggi} 
\affiliation{\UNIRM}
\author {P.~Valente} 
\affiliation{\INFNRM}

\date{\today}
\begin{abstract}
High energy positron annihilation is a viable mechanism to produce dark photons ($\Apr$). This reaction plays a significant role in beam-dump experiments using experiments using multi-GeV electron-beams on thick targets by enhancing the sensitivity to $\Apr$
production. The positrons produced by the electromagnetic shower can produce an  $\Apr$ via non-resonant ($e^+ + e^- \to \gamma + \Apr$) and resonant  ($e^+ + e^- \to \Apr$) annihilation on  atomic electrons. For {\it visible } decays, the contribution of resonant annihilation results in a larger sensitivity with respect to limits derived by the commonly used $\Apr$-strahlung in certain kinematic regions. When included in the evaluation of the E137 beam-dump experiment reach, positron annihilation pushes the current limit on $\varepsilon$ downwards by a factor of two in the range 33 MeV/c$^2<m_\Apr<120$ MeV/c$^2$.
\end{abstract}
\maketitle
\section{Motivations}
Fitting dark matter (DM) in the Standard Model (SM) of elementary particles is one of the most prominent open questions of contemporary physics. Null results in direct detection of halo DM calls for alternative explanations to the current WIMPs paradigm \cite{Arcadi:2017kky}. One of them conjectures the existence of a new class of lighter elementary particles not charged under the SM strong, weak, or electromagnetic forces.
A well motivated scenario considers DM with mass below 1 GeV/c$^2$, charged under a new $U(1)_D$ gauge symmetry, that interacts with the SM particles via the exchange of a light spin-1 boson (a {\it heavy photon} or $\Apr$, also called  {\it dark photon}). The coupling between SM particles and dark photons is induced by the kinetic mixing operator. This mechanism, originally  suggested by Holdom~\cite{HOLDOM1986196} as a possible minimal extension of the SM, has been lately interpreted as a {\it portal} between the SM world and a new {\it Dark Sector}~\cite{PhysRevD.88.114015,PhysRevD.80.095024}.
The low energy effective Lagrangian extending the SM to include dark photons can be written as follows:
\begin{equation}
\mathcal{L}_{eff} = \mathcal{L}_{SM} -\frac{1}{4} F'_{\mu\nu}F'^{\mu\nu} 
+ \frac{1}{2} m^2_{\Apr} \Apr_{\mu}{\Apr}^{\mu} 
- \frac{\varepsilon}{2} F'_{\mu\nu}F^{\mu\nu},
\end{equation}
where $F'_{\mu\nu}$ is the field strength of the hidden gauge field  $\Apr_{\mu}$, $m_{\Apr}$ the mass of the heavy photon, and $F_{\mu\nu}$ the SM photon field strength. The kinetic mixing parameter $\varepsilon$ is expected to be small, in the range of  $\sim 10^{-4} - 10^{-2}$~($\sim 10^{-6} - 10^{-3}$) if the mixing is generated by one (two)-loops interaction~\cite{HOLDOM1986196,Essig:2010ye,DELAGUILA1988633,ArkaniHamed:2008qp}. 
Depending on the relative mass of the $A'$ and the DM particles, the $\Apr$ can decay to SM particles ({\it visible } decay) or to light DM states ({\it invisible } decay).

This new idea generated many theoretical and phenomenological studies~\cite{BOEHM2004219,PhysRevD.90.115005,PhysRevD.80.075018,PhysRevD.88.114015,PhysRevD.80.095024,PhysRevD.91.094026,PhysRevLett.115.251301,PhysRevD.86.035022,PhysRevD.90.115014}, stimulated the reanalysis and interpretation  of old data~\cite{PhysRevLett.113.171802,PhysRevD.84.075020,BABAR,BABAR2,KLOE,WASA}, and promoted new experimental programs, searching both for the $\Apr$~\cite{PhysRevLett.106.251802,PhysRevLett.107.191804,CORLISS2017125,HPS,PADME,VEPP3,BE8,CORNELL} and for light DM states~\cite{PhysRevLett.118.221803,BDX,LDMX,PhysRevD.92.095014,PhysRevD.91.055006,PhysRevLett.118.011802}. 
For a comprehensive review of the subject, we refer the reader to Ref.~\cite{Alexander:2016aln,Battaglieri:2017aum}.


%
%

\begin{figure}
\vspace{1.cm} 
\includegraphics[width=.45\textwidth]{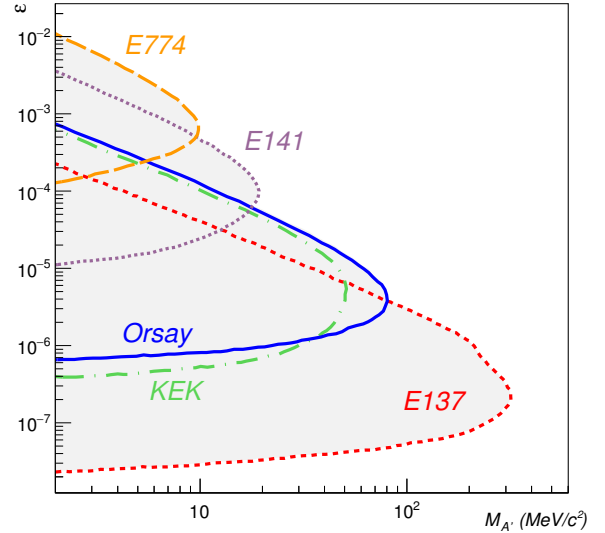}
\caption{\label{fig-limits} Limits on coupling $\varepsilon$ vs. $\Apr$ mass resulting from null results reported in beam-dump experiments, assuming dark photons decay visibly. Curves report limits derived in Ref.~\cite{PhysRevD.86.095019}.}
\end{figure}

In this context, accelerator-based experiments that make use of an intense electron-beam of moderate energy ($\sim$ 10 GeV) dumped on a thick target (beam-dump) are sensitive to  a wide area of  $\varepsilon$ vs. $m_{\Apr}$ parameter space~\cite{PhysRevD.80.075018,PhysRevD.86.095019}. 
Figure \ref{fig-limits} shows limits in ($\varepsilon$ vs. $m_{\Apr}$) extracted from the re-analysis of past electron beam-dump experiments data~\cite{PhysRevLett.59.755,PhysRevD.38.3375,PhysRevLett.67.2942,PhysRevLett.57.659,DAVIER1989150}, assuming the $\Apr$ decays visibly~\cite{PhysRevD.86.095019,PhysRevD.96.016004}. 
Even if, in principle, the concept is the same, the limited intensity prevented so far to run beam-dump experiments that  use a primary positron beam. 
{In this paper we show that positrons play an important role also in electron beam-dump experiments, where they are copiously produced in the electromagnetic shower developing in the thick target.}
When the sizable contribution of  secondary positron annihilations via non resonant and resonant production is considered, exclusion limits are pushed downwards in certain kinematics region, improving the experimental sensitivity of electron beam-dump experiments.\\
The paper is organized as follows. In Section~\ref{sec:production} we briefly discuss the $\Apr$ production mechanism by $O$(GeV) positrons impinging on a fixed target. In Section~\ref{sec:electronBeam} we focus on the electron-beam thick-target case. After reviewing the main features of positrons production in an electron-induced electromagnetic shower, we discuss the effect of the resulting $e^+$ angular and energy distribution on $\Apr$ production in the dump and its detection in a distant detector.
Finally, Section~\ref{sec:E137} presents the application of this new approach to the E137 experiment~\cite{PhysRevD.38.3375}, the electron-beam fixed target effort that placed the most stringent exclusion limits in the $\Apr$ parameters space.

\section{$\Apr$ production by positrons}\label{sec:production}
\begin{figure}[t!]
\begin{center}
\includegraphics[scale=0.65]{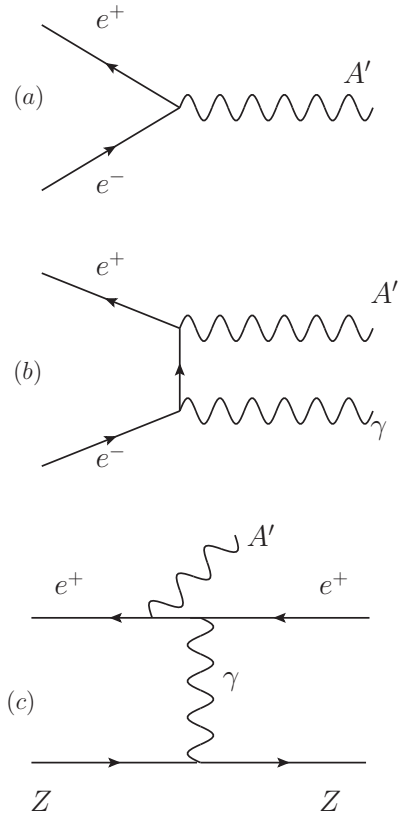}
\caption{Three different $\Apr$ production mechanisms by high-energy positrons on a fixed target: 
  $(a)$~resonant $\Apr$ production in $e^+e^-$ annihilation;
  $(b)$~$\Apr$-strahlung in $e^+e^-$ annihilation; 
  $(c)$~$\Apr$-strahlung in $e^+$-nucleus scattering.}
\label{fig-mech}
 \end{center}
\end{figure}

The processes involved in dark photon production by $O$(GeV) positrons are shown in Fig.~\ref{fig-mech}.
Diagram $(a)$ and $(b)$ describe, respectively, the $\Apr$ production through resonant ($e^+ + e^- \to \Apr$) and non-resonant ($e^+ + e^- \to \gamma +  \Apr$) positrons annihilation on atomic electrons. The first production mechanism was recently put forth~\cite{NewDirections,PADMEeplus} as a powerful tool to test at the forthcoming PADME experiment~\cite{PADME} hints of a 17~MeV $\Apr$ from anomalous $e^+e^-$ production in $^8$Be nuclear transitions~\cite{Krasznahorkay:2015iga}. 
Diagram c), instead, represents the ``$\Apr$-strahlung'' process, i.e. the radiative $\Apr$ emission by an impinging $e^+$ in the EM field of a target nucleus. At first order, the corresponding cross-section is the same as the one for the equivalent $e^-$ process \cite{Tsai:1973py}.

The cross-sections for the annihilation processes scale as $\varepsilon^2 \alpha$ (resonant) and $\varepsilon^2 \alpha^2$ (non-resonant), where $\alpha$ is the electromagnetic fine-structure constant, compared to the $\varepsilon^2 \alpha^3$ dependence characteristics of the $\Apr$-strahlung diagram. Specifically, the resonant diagram in Fig.~\ref{fig-mech}$a$ yields the total cross section: 
\begin{equation}
\sigma_{r} = \sigma_{\rm peak} \,\frac{\Gamma^2_\Apr/4}{(\sqrt{s}- m_\Apr)^2+\Gamma_\Apr^2/4}\; ,
\end{equation}
where $m_e$ is the electron mass, $s$ is the $e^+$ $e^-$ system invariant mass squared, $\sigma_{\rm peak}= 12 \pi/m_\Apr^2$ is the resonant cross section at the peak, and $\Gamma_\Apr = \frac{1}{3}m_\Apr \varepsilon^2 \alpha$ is the $\Apr$ decay width in the limit $m_e/m_\Apr \to 0$. Given that $\Gamma_\Apr/m_\Apr \ll 1$ if $\varepsilon\ll 1$,
for the resonant case the narrow-width approximation~\cite{PDG} has been consistently  used. The differential and total cross-sections for the non-resonant diagram in Fig.~\ref{fig-mech}$b$ read:

\begin{align}\label{eq:sigres}
\frac{d\sigma}{dz} =& \frac{4\pi\varepsilon^2\alpha^2}{s}\left(
\frac{s-m^2_\Apr}{2s}\frac{1+z^2}{1-\beta^2z^2}+\frac{2m^2_\Apr}{s-m^2_\Apr}\frac{1}{1-\beta^2 z^2}\right)\\
\sigma_{nr}=& \frac{8\pi\alpha^2 \varepsilon^2}{s}\left[ \left( \frac{s-m^2_\Apr}{2s}+\frac{m^2_\Apr}{s-m^2_\Apr}\right)\log{\frac{s}{m^2_e}} - \frac{s-m^2_\Apr}{2s} \right] \label{eq:sigres2}\;,
\end{align}
with $z$ being the cosine of the $\Apr$ emission angle in the $e^+e^-$ rest frame, measured with respect to the positron axis, and $\beta=\sqrt{1-\frac{4m_e^2}{s}}$. It is worth mentioning that these results were derived at tree level, keeping the leading $m_e$ dependence to avoid non-physical divergences when $\left | z \right| \rightarrow 1$ \cite{VEPP3}. 
To avoid infrared divergences when $s\rightarrow m^2_\Apr$, we applied a low-energy cut-off for the non-resonant mode. We required that the real photon energy in the center-of-mass frame is at least $1\%$ of the impinging positron energy. This cut-off value is commonly adopted in such calculations~ \cite{Bohm:1983rn}. The cut-off discriminates between the ``hard'' regime, where Eq.~\ref{eq:sigres2} is applicable, and the ``soft'' one. The low-energy contribution to the total cross-section should be re-absorbed in the resonant part, resulting in an effective enlargement of the $\Apr$ width. For our specific case, the enlargement is $\simeq 20\%$. This affects the signal yield $Y$ in case of $\Apr$ resonant production, directly proportional to the resonance width. However, as discussed in the following Section, the dependence of the exclusion limit on $Y$ is weak. This makes the correction negligible.


%

\begin{figure}[t!]
\begin{center}
\includegraphics[scale=0.45]{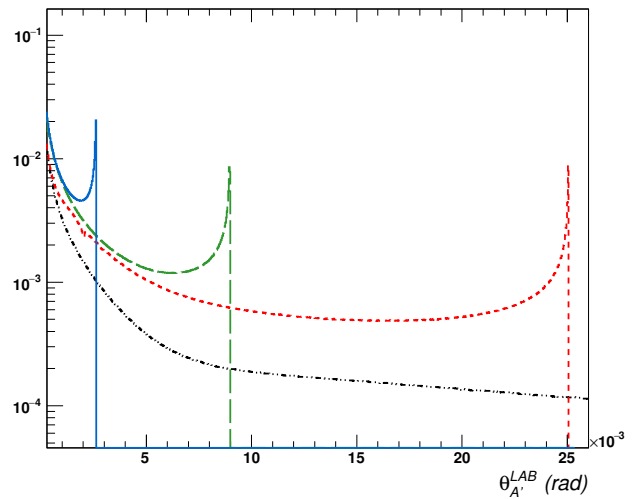}
\caption{The angular distribution in the laboratory frame of $\Apr$ produced in non-resonant annihilation of 20 GeV positrons for different dark photon masses: 20 MeV/c$^2$ (red, short-dashed), 50 MeV/c$^2$ (green, long-dashed), 100 MeV/c$^2$ (blue, continuous). For comparison, the angular distribution of photons in the process $e^+e^- \rightarrow \gamma \gamma$ is also shown (black, dash-dotted).}
\label{fig:AprAngle}
 \end{center}
\end{figure}

The main kinematic characteristics of the two annihilation mechanisms are as follows. In case of resonant positrons annihilation, the kinematics of the produced $\Apr$ is strongly constrained by the one-body nature of the final state. A dark photon with mass $m_\Apr$ is produced with energy $E_R=\frac{m^2_\Apr}{2m_e}$,  in the same direction of the impinging positron.
For the non-resonant case, instead, the $\Apr$ angular distribution in the CM frame, given by Eq.~\ref{eq:sigres}, is  concentrated in the $e^+ e^-$ direction, due to the $1-\beta^2 z^2$ factor at denominator. This results in an angular distribution in the laboratory frame strongly peaked in the forward direction, the effect being more intense for larger values of the $\Apr$ mass. The maximum $\Apr$ emission angle in the laboratory frame is $\theta^{Max}_\Apr \simeq \frac{s-m^2_{\Apr}}{2 m_\Apr E_0}$ (see Fig.~\ref{fig:AprAngle}). The corresponding energy distribution ranges from $E_R$ to the primary positron energy $E_0$, with an average value of $\frac{E_0}{2}(1+\frac{m^2_\Apr}{2m_e E_0})$. 

\section{Electron-beam on a thick-target}\label{sec:electronBeam}


In the following, we investigate the possibility of using the $\Apr$ production processes by positrons previously described in an electron-beam fixed-target scenario, exploiting secondary $e^+$ emitted through standard electromagnetic processes.
When a high-energy electron impinges on a material, it initiates an “electromagnetic shower” (EM), that is a cascade multi-particle production process. The two main reactions contributing to the process are photons production through bremmstrahlung by electrons and positrons, and $e^+ e^-$ pairs production by photons. 
As a consequence, after few radiation lengths the developing shower is made by an admixture of electrons, positrons and photons, characterized by different energy distributions. 

In previous papers describing $\Apr$ production in electron beam-dump experiments~\cite{PhysRevD.86.095019,PhysRevD.86.095019} only the bremmstrahlung-like dark photon production  by electrons has been included (we refer to Ref.~\cite{PhysRevD.95.036010} for a critical discussion of limitations of the widely used Weizs\"acker-Williams approximation within this context). 
In this work, we discuss how, in the positron-rich environment produced by the high-energy electron showering in a beam-dump, contributions from non-resonant and resonant annihilation can be sizable. As a rule of thumb, compared to the characteristic shape of $\Apr$-strahlung exclusion limits, the resonant annihilation process leads to a more stringent constraint at low $\varepsilon$, in the $\Apr$ mass window constrained by the primary beam energy (right bound) and by the detection threshold (left bound).

To evaluate the $\Apr$ production by positrons and the subsequent detection of the visible decay products ($e^+ e^-$ pairs) in an electron-beam dump experiment we employed the following Montecarlo procedure. First, we evaluated the energy spectrum and the multiplicity of secondary positrons in the beam-dump through a Geant4 based~\cite{AGOSTINELLI2003250} simulation. Then, we used this result as input for a custom Montecarlo code that generates $\Apr$ events according to the two positron annihilation processes described above, handles the $\Apr$ propagation and subsequent decay to an $e^+e^-$ pair, and includes the experimental acceptance of a detector placed downstream of  the dump. The code also computes the total number of produced particles per electron-on-target (EOT).
We found this dual-step method more effective than including the dark photon as a new particle in Geant4, together with the corresponding production mechanisms. Decoupling the $\Apr$ production in the dump from the EM shower development allows to handle cases with $\varepsilon \ll 1$ without generating a very large number of EM showers in the target. This is possible by artificially enhancing the $\Apr$ production cross section in the second step, properly accounting for this in the final reach evaluation. 
Also, this method saves computation time while evaluating the sensitivity of a beam-dump experiment as a function of the $\Apr$ mass and coupling. Indeed, for a given experimental setup, the first step is performed only once, while only the second, i.e. $\Apr$ generation  and  detector acceptance evaluation, is repeated for different dark photon masses and couplings. 

\subsection{$\Apr$ production yield}

The total $\Apr$ yield per EOT due to positrons annihilation is given by:
\begin{equation}\label{eq-prod}
N_{A^\prime}= \frac{N_A}{A} Z \rho \int_{E^R_{min}}^{E_0} dE_e \,\, T_+(E_e)\,\sigma(E_e) \; ,
\end{equation}
where $A$, $Z$, $\rho$, are, respectively, the target material atomic mass, atomic number, and mass density, $E_0$ is the primary beam energy, $N_A$ is Avogadro's number, $\sigma(E_e)$ is the energy-dependent $\Apr$ production cross-section, and $E^R_{min}=\frac{m^2_\Apr}{2m_e}$ is the minimal positron energy required to produce a dark photon with mass $m_{\Apr}$ through positron annihilation on atomic electrons (see Section~\ref{sec:production}). Finally, $T_+(E_e)$ is the positrons differential track-length distribution.  We note that the same approach applies in case of bremsstrahlung-like $\Apr$ production by electrons, with  $\sigma(E_e)$ replaced by the corresponding cross-section and the track-length being that of electrons in the dump.

Since the typical dark photon width accessible by beam dump experiments is much smaller than the scale of $T_+$ variations, Eq.~(\ref{eq-prod}) reduces to:

\begin{equation}
\label{eq:NAp}
N_{A^\prime} \simeq \frac{\pi}{2} \frac{N_A}{A} Z \rho \, \sigma_{\rm peak}\Gamma_\Apr \, \frac{m_\Apr}{m_e} \,  T_+(E_R) \; 
\end{equation}
in case of $\Apr$ resonant production.


\subsection{Positrons track-length distribution}

The positrons track-length distribution $T_+(E_e)$ is
defined as the integral over the beam-dump volume of the differential fluence $\Phi(E_e)$, corresponding to the density of particle tracks in the volume \cite{Chilton}. Intuitively, the quantity $T_+(E_e)dE_e$ represents the total path length in the dump taken by positrons with energy in the interval between $E_e$ and $E_e+dE_e$.

The differential track-length can be calculated by assuming that all particles propagate along the primary beam axis, thus neglecting the transverse contribution to the path length from angular straggling. This approximation is well justified by the fact that the electromagnetic shower is strongly peaked in the forward direction, so that the corresponding effects on the shape and the normalization of $T_{+}(E_e)$ are negligible. We underline that the longitudinal approximation is valid in the context of $T_+(E_e)$ calculation, while angular effects in the shower have to be considered when computing the detection acceptance, as described later.

Under this hypothesis, $T_+(E_e)$ can be obtained by integrating the differential energy distribution $I^+_e(E_e,t)$ of positrons over the full beam-dump length: 

\begin{equation}\label{eq:track-length}
T_+ (E)=\int_{0}^{L_{Dump}} I^+_e(E_e,t) dt \; .
\end{equation}

Here, $I^+_e(E_e,t)$ is normalized so that $\int_0^{E_0}I^+_e(E_e,t)dE_e$ is the total positrons current per EOT through a plane perpendicular to the beam axis, located at the depth $t$ in the beam-dump.

The Geant4-based application we developed to evaluate $T_+(E_e)$ for a generic electron thick-target setup works as follows. The target length $L_{Dump}$ is divided in $N$ thin layers of thickness $\Delta t$, located at $t_i$, and the differential positrons current $I_e^+(E)$, normalized per EOT, is sampled on a plane positioned at depth $t_i$. The differential track-length $T_+(E)$ has then been obtained summing over the different planes, and multiplying by the layer thickness:

\begin{equation}\label{eq:track-lengthMC}
T_+(E_e) = \Delta t \sum_{i=1}^{N} I^e_+(E_e,t_i) \; .
\end{equation}

The values of $N$ and $\Delta t$ have to be tuned to the primary electron-beam energy and to the beam-dump characteristics. For a multi-GeV electron-beam impinging on a thick ($L_{dump}~\gg~X_0$) target, this procedure yields stable results for $\Delta t < X_0/10$ and $N>200$. We also verified that no appreciable variations in results are found when, in the previous equation, the differential fluence $\Phi^e_+(E_e,t_i)$ of positrons sampled at the depth $t_i$ is used, thus confirming the validity of the longitudinal approximation. 

\begin{figure}
\vspace{1.cm} 
\includegraphics[width=.5\textwidth]{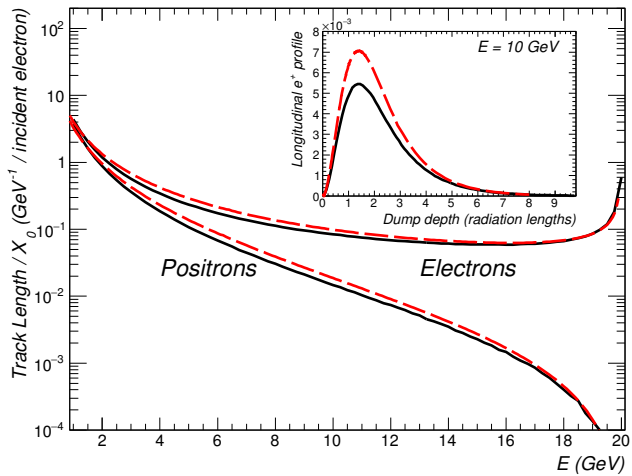}
\caption{\label{fig-bd-spectra} Track-length distribution $T(E_e)$ for a 20-GeV electron-beam impinging on an Aluminum beam-dump, divided by the corresponding radiation length. The red dashed line corresponds to the prediction of Tsai~\cite{PhysRev.149.1248}. The inset shows the longitudinal positrons shower profile at $E_e=10$ GeV (in GeV$^{-1}$ per incident electron). The red dashed line corresponds to the prediction of Tsai parametrization.}
\end{figure}

As an example, Fig.~\ref{fig-bd-spectra} shows the differential track-length distribution of positrons and electrons, for a 20-GeV electron-beam impinging on an Aluminum beam-dump. It is worth noticing that the two distributions have different shape at high energy since
positrons are generated only as secondary particles in the electron-induced process.  To validate the result, we also compared it with the output of a FLUKA-based simulation~\cite{BOHLEN2014211,Ferrari:2005zk} where the built-in differential track-length scorer has been employed, finding an agreement at a few $\%$ level in the full energy range. 
In the following, we considered this difference as the systematic error associated to $T_+(E_e)$. Since, at a fixed $M_\Apr$, limits on $\varepsilon$ scale roughly as $(T_+)^{\frac{1}{4}}$ (lower bound) or $\log(T_+$) (upper bound), the corresponding effect on the result is negligible.

The prediction of the analytical model of Ref.~\cite{PhysRev.149.1248} is shown as a red dashed line in Fig.~\ref{fig-bd-spectra} (in the calculation, we considered contributions up to second generation electrons and positrons). The model well reproduces the electrons and positrons track-length distribution in the full energy range.
We also note that the model remarkably matches the shape of the longitudinal shower of both particles, i.e. the dependence of $I^e_\pm(E_e,t)$ as a function of the depth $t$ for fixed positrons energy $E_e$, describing the shower evolution in the dump. The inset in the same figure shows the specific case of $E_e=10$ GeV positrons, with $I^+_e(E_e,t)$ normalized to the Aluminum radiation length. 
\subsection{Angular effects}

\begin{figure}
\vspace{1.cm} 
\includegraphics[width=.5\textwidth]{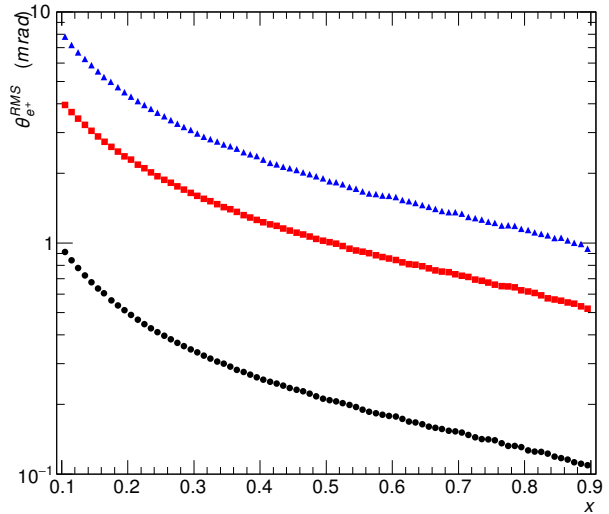}
\caption{\label{fig-bd-angles} RMS value of the differential positrons track-length angular distribution $T_+(E_e,\vec{\Omega}_e)$, as a function of $x=E/E_0$, for different values of the primary electron-beam energy $E_0$: $100$ GeV (black circles), $20$ GeV (red squares), $11$ GeV (blue triangles).}
\end{figure}

The angular spread of positrons in the shower has a non-negligible effect in dark photon production and detection. It induces a further widening in the $\Apr$ angular distribution that sums up to to the intrinsic spread due to the production mechanism and to the $e^+e^-$ decay. The latter is characterized by an average opening angle between the two leptons $\theta_D^{e^+e^-} \simeq \frac{M_\Apr}{E_\Apr}$. This may result in a sizable fraction of particles produced out of the detector geometrical acceptance.

The double-differential positrons track length distribution $T_+(E_e,\vec{\Omega}_e)$ is required in order to account for this effect, with the momentum direction $\vec{\Omega}_e$ measured with respect to the primary beam axis. To evaluate it, we used the previously described Montecarlo-based procedure, replacing the quantity $I^e_+$ in Eq.~\ref{eq:track-lengthMC} with the double-differential positrons current $I^e_+(E_e,\vec{\Omega}_e)$. This approach, even if approximated, is motivated by the lack in literature of a full analytical treatment of the secondary particles angular distribution in an EM shower developing in a thick target. Moreover, common particles transport codes, such as Geant4 and FLUKA, although containing built-in scorers to estimate the track length of a specific particle species within a volume, only provide results integrated over the full solid angle. It is worth noticing that, having the same dependence of the track-length distribution, exclusion limits are weakly affected by any uncertainty on angular distributions.

Figure~\ref{fig-bd-angles} shows the RMS value of the differential positrons track-length angular distribution, as a function of $x=E_e/E_0$, for different values of the primary beam energy. The results scale approximately as $1/E_e$. This reflects the energy dependence of the two main contributions to positrons angular distribution in the dump: the bremsstrahlung and pair production characteristic emission angle $\theta_1 \propto \frac{m_e}{E_e}$, and the multiple-scattering spread $\theta_2 \propto \frac{E_s}{E_e}$, with $E_s = m_e \sqrt{4\pi / \alpha} \simeq 21.2$ MeV.

\begin{figure*} 
\includegraphics[width=.9\textwidth]{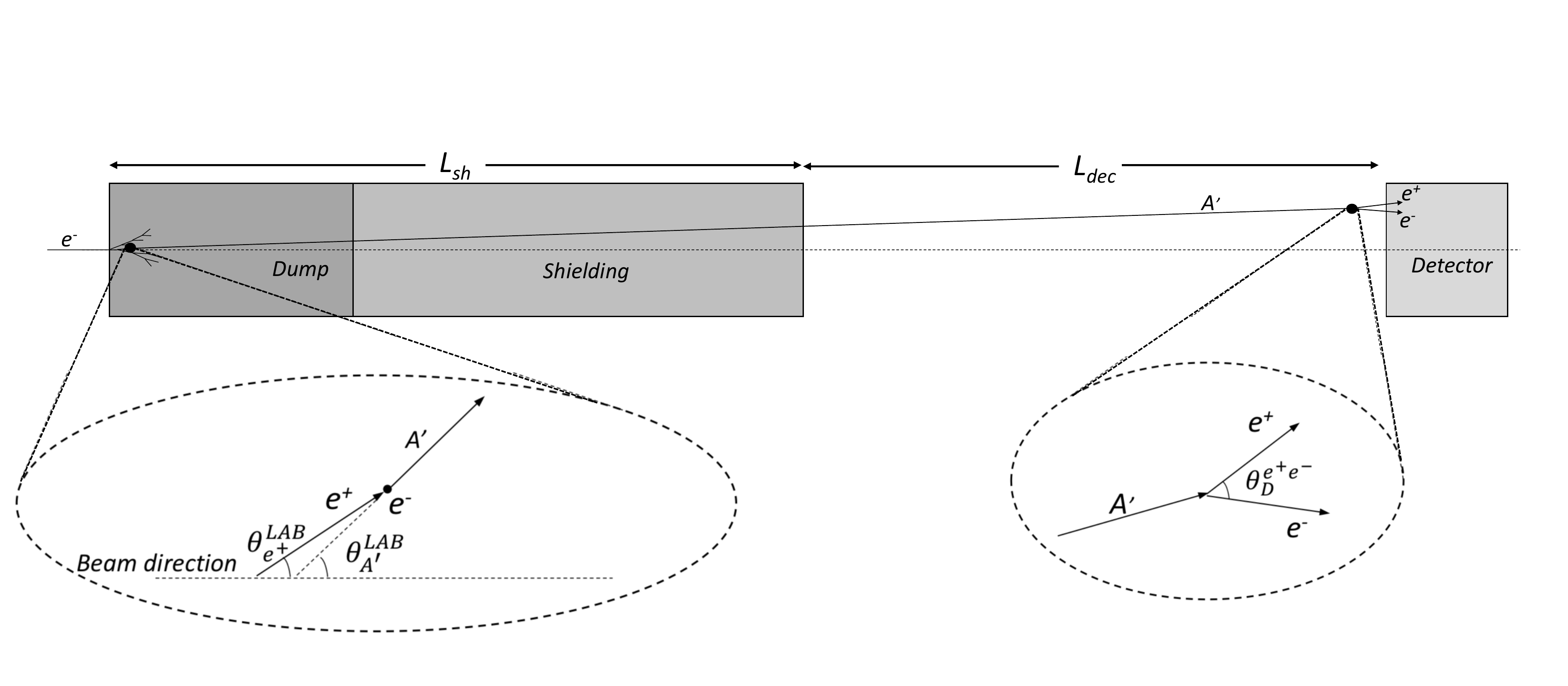}
\caption{\label{fig-decayScheme} Typical setup of a beam-dump experiment for visible decay $\Apr$ search. $L_{sh}$ is the total length of target and shielding, while $L_{dec}$ is the length of the downstream decay region, preceding the detector. The two insets show schematically the angles involved in the $\Apr$ production and decay processes, as described in the text.}
\end{figure*}

A proper evaluation of angular effects is particularly critical in case of resonant $\Apr$ production.  In this process, the dark photon is always produced in the positron direction.  Therefore, $\Apr$s and positrons have the same angular distribution. For example, in case of $M_\Apr = 50$ MeV/c$^2$ and a primary electron-beam with $E_0=20$ GeV, dark photons, produced by positrons with energy $E_R \simeq 2.45 $ GeV, would be emitted with an angular spread of about 4 mrad (see Fig.~\ref{fig-bd-angles}), comparable to the opening angle of the $e^+ e^-$ decay pair of about 20 mrad.


%

\subsection{Total signal yield in the detector}

A general treatment of the total signal yield in a distant on-axis detector in a beam-dump experiment searching for visible dark photon decay is reported in Ref.~\cite{PhysRevD.86.095019}. Here, we briefly summarize the results, explicitly including the angular dependence of the produced dark photons. Since the typical distance between the detector and the beam dump is much larger than the length of the latter, to compute the detection acceptance we neglect the longitudinal dependence of the $\Apr$ production vertex, fixed at $t=0$ (see Eq.~11 and related comments in the aforementioned reference).  

The differential dark photon distribution per EOT, reads (see Eq.~\ref{eq:NAp} for target luminosity factors):
\begin{eqnarray}\label{eq:apr-distr}
\frac{dN}{dE_\Apr \,d\vec{\Omega}_\Apr} & \propto & \int _{E^R_{min}}^{E_0} dE_e \int_{4 \pi} d\vec{\Omega}_e T_+(E_e,\vec{\Omega}_e)
\nonumber \\*
&& \frac{d\sigma(E_e)} {d\vec{\Omega}^\prime_\Apr}\,\delta(E_\Apr - f(E_e,\vec{\Omega}^\prime_\Apr)) \;  ,
\end{eqnarray}
where $\frac{d\sigma(E_e)} {d\vec{\Omega}^\prime_\Apr}$ is the differential $\Apr$ production cross section, $\vec{\Omega}_\Apr$  and $\vec{\Omega}^\prime_\Apr$ are, respectively the dark photon momentum direction in the laboratory frame and in the rotated positron frame, and $f(E_e,\vec{\Omega}^\prime_\Apr)$ is the kinematic function relating the $\Apr$ energy to the  impinging positron energy and to the $\Apr$ emission angle.


After being produced, dark photons propagate along the direction $\vec{\Omega}_\Apr$ with a differential decay probability per unit path given by: 
\begin{equation}
\frac{dP}{dl}=\frac{1}{\lambda}e^{-l/\lambda} \; ,
\end{equation}
where $\lambda = \frac{E_\Apr}{m_\Apr}\frac{1}{\Gamma_\Apr}$ is the $\Apr$ decay length.

Electron and positrons from the $\Apr$ decay are emitted on a cone with typical aperture $\theta^{e^+e^-}_D\simeq \frac{M_\Apr}{E_\Apr}$ with respect to the  $\vec{\Omega}_\Apr$ axis. The total signal yield is thus obtained combining the $\Apr$ angular distribution (Eq.~\ref{eq:apr-distr}) with the decay kinematics, and integrating the result over the geometrical acceptance of the detector. The latter can be roughly determined as the product of a longitudinal factor $\varepsilon_L$ depending on the shielding $L_{sh}$ and decay region $L_{dec}$ length and a transverse factor $\varepsilon_T$ related to the detector face width $S$ (see Fig.~\ref{fig-decayScheme}):
\begin{eqnarray}
\varepsilon_L & \simeq & e^{-L_{sh}/\lambda} \cdot (1-e^{-L_{dec}/\lambda}) \\*
\varepsilon_T & \simeq & S \,/\, (\theta^{RMS}_\Apr (L_{sh}+L_{dec}) \oplus \theta^{e^+e^-}_D L_{dec}) \; ,
\end{eqnarray}
with $\theta^{RMS}_\Apr$ being the width of the dark photon angular distribution and $\theta^{e^+e^-}_D$ the typical opening angle between the $e^+e^-$ pair from $\Apr$ decay. It is worth noticing that the above expression for $\varepsilon_T$ holds exactly in case of $\Apr$ decay happening at the beginning of the decay volume (large $\varepsilon$ case). In other cases, it leads to a detection acceptance underestimate, since the contribution of $\theta_D^{e^+e^-}$ to the transverse displacement is actually smaller. Given that, for typical beam-dump experiments, $L_{dec} \approx L_{sh}$ and that $\theta^{RMS}_\Apr \approx \theta_D^{e^+e^-}$, the obtained result is valid within a factor of $\simeq$ 2.

To evaluate the detection acceptance, we generate a large set of $\Apr$ events by randomly randomly sampling positrons from the $T_+(E,\vec{\Omega}_e)$ distribution. For each positron, a dark photon is generated according to the production cross-section. Finally, the $\Apr$ is propagated along the $\vec{\Omega}_\Apr$ direction, with the $e^+e^-$ pair generated at the decay vertex assuming an isotropic distribution in the dark photon rest frame. The detector acceptance is determined vy counting the number of electrons/positrons hitting the detector. To speed-up calculations, dark photons are always forced to decay in the region between the shielding and the detector. A weight $\varepsilon_L$  is associated to each event to account for it.




\section{Exclusion limits from the E137 experiment}\label{sec:E137}

In this section we derive the contributions of  resonant and non-resonant $e^+$ annihilation  in the  specific case of the SLAC E137 experiment~\cite{PhysRevD.38.3375}. Among the past electron beam-dump experiments re-analyzed in the contest of $\Apr$ search
~\cite{PhysRevD.86.095019}, E137 is the one sensitive to the smallest values of $\varepsilon$, down to $\simeq 10^{-8}$ (see Fig.~\ref{fig-limits}).
We decided to focus on this experiment because, as shown below, the two new production mechanisms extend the exclusion limits to lower values of $\Apr$ coupling with respect to the bremsstrahlung-like diagram only.

The E137 experiment searched for long-lived neutral objects produced in the electromagnetic shower initiated by 20 GeV electrons in the SLAC beam-dump.
Particles produced in the water-cooled aluminum plates forming the dump, would have to penetrate 179 m of earth shielding and decay in the 204 m region downstream of the shield. The E137 detector consists of an 8-radiation length electromagnetic calorimeter made by a sandwich of plastic scintillator paddles and iron (or aluminum) converters. Multi-wire proportional chambers provided an accurate angular resolution, essential to keep the cosmic background to a negligible level. A total charge of $\sim$ 30~C was dumped during the live-time of the experiment in two slightly different experimental setups: in the first run (accumulated charge $\simeq$~10~C), the detector had a transverse size of 2$\times$3 m$^2$, while in the second run this was 3$\times$3 m$^2$.  

The original data analysis searched for axion-like particles decaying in e$^+$ e$^-$ pairs, requiring a deposited energy in the calorimeter larger than 1 GeV with a track pointing to the beam-dump production vertex. The absence of any signal provided stringent limits on axions/photinos.
Negative results were used in~\cite{PhysRevD.86.095019} and~\cite{PhysRevD.96.016004} to set strong constraints on to the visible decay $A' \to e^+ e^-$ assuming the $A^\prime$-strahlung (Fig~\ref{fig-mech}$a$) as the only production mechanism. Including the resonant and non-resonant positron annihilation, we have derived extended and more accurate limits for the $\Apr$ coupling to SM particles.

{\centering
\begin{figure}
\vspace{1.cm} 
\includegraphics[width=.53\textwidth]{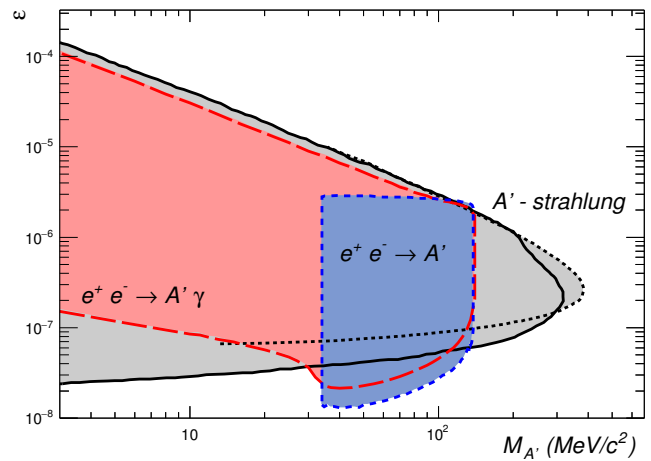}
\caption{\label{fig-e137}
Exclusion limits on the $\varepsilon$ vs $M_\Apr$ parameter space derived from the E137 experiment considering considering $e^+$ non-resonant (long-dashed red line) and resonant (short-dashed blue line) production. Results from previous analysis which included only production via $\Apr$-strahlung  are depicted as black-solid~\cite{PhysRevD.86.095019} and black-dotted~\cite{PhysRevD.96.016004} lines. }
\end{figure}
}

To derive the E137 exclusion limits for resonant and non-resonant $\Apr$, we used the Montecarlo-based numerical approach described above. The experimental acceptance was evaluated separately for the two E137 runs and combined with proper weights to account for the different accumulated charges. In the calculation, we employed the same selection cuts used in the original analysis:
\begin{itemize}
\item The energy of the impinging $e^+$/$e^-$ particle has to be larger than 1 GeV. We note that, in case of resonant production, this put a hard limit on the minimum value of the $\Apr$ mass of about 33 MeV/c$^2$.
\item The angle of the impinging particle on the detector, measured with respect to the primary beam axis, has to be smaller than 30 mrad.
\end{itemize}
We found that both particles from $\Apr$ decay hit the detector in a non-negligible fraction of events. In these cases, we applied previous selection cuts respectively considering the sum of the two energies to be greater than 1 GeV and the energy-averaged impinging angle to be less than 30 mrad.

Based on the null observation  reported by E137, we derived the exclusion contour considering a $95\%$ C.L. upper limit of 3 events.
Figure~\ref{fig-e137} shows results for both  resonant (short-dashed blue line) and non-resonant (long-dashed red line) annihilation. Limits  obtained by the $\Apr$-strahlung from Ref.~\cite{PhysRevD.86.095019} and ~\cite{PhysRevD.96.016004} are shown in the figure as a black solid line and a black dotted line, respectively.  Resonant annihilation provides the best exclusion limits for  $m_{\Apr}$ in the (33 MeV/c$^2<m_\Apr<120$ MeV/c$^2$) range, strengthening  by  almost a factor of two the previous limits. The lowest limit on $\varepsilon \sim 10^{-8}$ is obtained for $m_{\Apr}=33$~MeV/c$^2$.  In case of resonant annihilation, the sharp cut-off at low mass is determined by the energy detection threshold. At large $\varepsilon$ the reach is limited by  the small $\Apr$ decay width, not sufficiently dilated by the Lorentz boost factor and thus resulting in the $\Apr$ decay within the shielding. The non-resonant contribution is slightly less sensitive but extends the reach to lower masses down to $m_{\Apr} \sim$ few  MeV/c$^2$, for $\varepsilon$ values ranging from $O$(1) to $O$(10) with respect to the limit obtained by considering the $\Apr$-strahlung.

\section{Conclusions}

In this paper, we showed that $e^+$ resonant and non-resonant annihilation are two viable dark photon production mechanisms competitive with the widely considered $\Apr$-strahlung. This argument can be applied to electromagnetic showers initiated by an electron-beam in a thick target. We used a Montecarlo-based approach to numerically derive the energy and angular distribution of $\Apr$ produced in a beam-dump and evaluated the effect on the accepted yields on a downstream detector. We explicitly recalculated the reach of E137 experiment showing that, taking into account resonant and non-resonant annihilation, the exclusion limits in the $m_\Apr$ range (33 MeV/c$^2$ - 120 MeV/$c^2$) are pushed down by a factor of two. This work shows that secondary positron annihilation needs to be included for a correct evaluation of all the exclusion limits obtained with electron-beams.

 \section*{Acknowledgments}
 C.D.R.C.~acknowledges financial support from 
 COLCIENCIAS (doctoral scholarship 727-2015).
 E.N.~is supported in part by the INFN ``Iniziativa
Specifica'' TAsP-LNF.

\bibliographystyle{apsrev4-1} 
\bibliography{biblio} 

\end{document}